\documentclass[aps,prl,twocolumn,showpacs,preprintnumbers,amsmath,amssymb,superscriptaddress]{revtex4}
\usepackage{graphicx}% Include figure files
\usepackage{dcolumn}% Align table columns on decimal point
\usepackage{bm}% bold math
\usepackage{ifthen}

\renewcommand{\vec}[1]{{\bf #1}}

\begin{document}

\title{Importance of second-order piezoelectric effects in zincblende semiconductors}

\author{Gabriel Bester}
\affiliation{National Renewable Energy Laboratory, Golden, Colorado 80401}
\author{Xifan Wu}
\author{David Vanderbilt}
\affiliation{Department of Physics and Astronomy, Rutgers University,
Piscataway, New Jersey 08854, USA}
\author{Alex Zunger}
\affiliation{National Renewable Energy Laboratory, Golden, Colorado 80401}
\email{gabriel_bester@nrel.gov}
\homepage{sst.nrel.gov}
\date{\today{}}% It is always \today, today,

\begin{abstract}
We show that the piezoelectric effect that describes
the emergence of an electric field in response to a crystal deformation
in III-V semiconductors such as GaAs and InAs 
has strong contributions from second-order effects that have been
neglected so far.  We calculate the second-order piezoelectric tensors
using density functional theory and obtain the piezoelectric field
for [111]-oriented In$_x$Ga$_{1-x}$As quantum wells of realistic
dimensions and concentration $x$. We find that the linear and the 
quadratic piezoelectric coefficients have the opposite effect on the field,
and for large strains the quadratic terms even dominate.
Thus, the piezoelectric field turns out to be a rare example of a
physical quantity for which the first- and second-order
contributions are of comparable magnitude.
\end{abstract}



% 77.65.-j Piezoelectricity and electromechanical effects
% 77.65.Bn Piezoelectric and electrostrictive constants
% 77.65.Ly Strain-induced piezoelectric fields
% 78.67.De Quantum wells
% 71.15.Dx Computational methodology (Brillouin zone sampling, iterative diagonalization, 
%          pseudopotential construction)% 
% 71.15.-m Methods of electronic structure calculations

\maketitle

% INTRODUCTION
Since the discovery of piezoelectricity in 1880 by the Curie
brothers \cite{Cady46}, widespread efforts have been aimed at 
understanding this peculiar effect and developing its
applications.   Piezoelectric materials are in use today in a wide
range of devices, including ultrasonic transducers
for medical and sonar imaging and various types of
micropositioners and actuators.
Since the early days, the effect has been understood as arising from
displacement of the ions in response to a mechanical deformation,
leading to the appearance of charges on some of the crystal's surfaces
\cite{Cady46}. Despite some early doubts, it is now well established
\cite{arlt68,martin72,hubner73,harrison74,king-smith93} 
that this is a {\it bulk} effect.  It has two components; the contribution
coming from the ionic displacements tends to be compensated by the
purely electronic (frozen-ion) response, resulting in a subtle balance
between ionic and electronic contributions
\cite{martin72,hubner73,harrison74}.

Until now, theoretical modeling of the piezoelectric effect in bulk 
solids \cite{degironcoli89},
quantum wells (Ref.~\onlinecite{smith90} and references
therein) and more recently in quantum dots
\cite{stier99,andreev00,holm02,sheng03b,ranjan03,bester05a} has focused
exclusively on the first-order piezoelectric tensor $\tilde{e}_{\mu j}$,
neglecting possible higher-order terms.
That is, if $P_\mu =
\sum_j e_{\mu j}\eta_j + \frac{1}{2}\sum_{jk}{B}_{\mu jk}\eta_j\eta_k
+ ...$, where $P$
is the polarization and $\eta$ is the strain, previous work has
concentrated on the linear coefficient $e_{\mu j}$ to the exclusion
of the quadratic coefficient $B_{\mu jk}$.
% Because of symmetry, these tensors can often be reduced to a few independent
% components, e.g., $e_{14}$ for zincblende material or $e_{31}$ and
% $e_{33}$ for wurzite materials. 

Recent experimental determinations of piezoelectric
constants (e.g., Refs.
\onlinecite{hogg93,sanchez-rojas94,bahder94,ballet98,ballet99,cho01,sanchez02})
have tended to follow this approach, interpreting the measured
piezoelectric fields by assuming a linear relationship between polarization
and strain (retaining $e_{\mu j}$
but neglecting $B_{\mu jk}$). Indeed,
the experimental procedures used thus far to deduce piezoelectric constants
from measured fields have made it easy to overlook 
the importance of the second-order piezoelectric effect,
because measurements were restricted to heterostructure quantum wells
with small lattice mismatch (those with large
lattice mismatch, like high-In-content
(In,Ga)As/GaAs structures, were avoided because they tend to develop
unwanted dislocations).
Similarly, experiments for quantum wells or bulk materials under
pressure have tended to probe only a very small region of strain, so
that a clear signature of the quadratic
dependence of field upon strain is difficult
to detect. 

In this Letter we show, using self-consistent density-functional
theory (DFT) calculations for GaAs and InAs, that the hitherto
neglected second-order piezoelectric tensor gives
significant contributions to the piezoelectric field.  We show that 
neglecting the second-order
piezoelectric tensor leads to an overestimation by 200\% in the
piezoelectric field for
In$_x$Ga$_{1-x}$As quantum wells on GaAs in the  experimentally
accessible range of concentration $x$.   For higher In concentrations,
accessible in
quantum dots, second-order terms will dominate over first-order
terms. This new insight is important because it represents a
paradigm shift in the interpretation of measurements of
piezoelectricity in quantum wells and quantum dots.

% Method

% Quantum wells or quantum dots with realistic
% sizes and compositions require large calculations with thousands of
% atoms \cite{bester05a}. To obtain the piezoelectric field for such large
% structures, we take the approach of first computing, once and
% for all, the piezoelectric response of each constituent bulk material
% as a function of strain from DFT and then use these parameters in 
% a classical force field approach, as subsequently outlined. 

We find it most convenient to formulate the piezoelectric response
in terms of the {\it reduced} (and dimensionless) polarization
$p_\mu$ defined implicitly via
$P_\alpha = \frac{e}{\Omega}\sum_\mu p_\mu a_\alpha^{(\mu)}$,
where $e$ is the charge quantum, $\Omega$ is the cell volume,
$P_\alpha$ is the polarization in Cartesian coordinates,
and $a_\alpha^{(\mu)}$ is the $\alpha$'th component of the
$\mu$'th strain-deformed lattice vector.
We expand this reduced polarization, retaining the {\it second-order} strain
as
\begin{equation}
\label{eq:red_pol}
p_\mu = \sum_j \tilde{e}^0_{\mu j}\; \eta_j +
\frac{1}{2}\sum_{jk}\widetilde{B}_{\mu jk}\; \eta_j\eta_k\quad .
\end{equation}
The reduced proper piezoelectric tensor is then
\begin{equation}
\tilde{e}_{\mu j} = \frac{dp_\mu}{d\eta_j} =
\tilde{e}^0_{\mu j} + \sum_{k}\widetilde{B}_{\mu jk}\, \eta_k
 \quad ,
\label{eq:estrain}
\end{equation}
where we use $\eta_j$ ($j$=1,6) to denote strain in the Voigt
notation.  Here $\tilde{e}^0_{\mu j}$ is the
reduced proper piezoelectric tensor of the unstrained material,
while $\widetilde{B}_{\mu jk}$ is a fifth rank tensor with Cartesian 
coordinates $\mu$ and the strain index in Voigt notation $j,k$
and represents the first-order change
of the reduced piezoelectric tensor with strain.
We obtain $\tilde{e}^0_{\mu j}$ and $\widetilde{B}_{\mu jk}$ from
first-principles calculations in a manner described next.

{\it First-principles calculation of linear and non-linear
piezoelectric coefficients:}
Symmetry considerations for the zincblende crystal
structure imply that the only nonzero elements of the piezoelectric
tensor are $\tilde{e}^0_{14}=\tilde{e}^0_{26}=\tilde{e}^0_{36}$
(i.e., there is only one independent element).  Similar considerations
guarantee that there are only 24 non-zero elements of the
$\widetilde{B}_{\mu jk}$ tensor, which can be reduced to three independent
elements $\widetilde{B}_{114}$, $\widetilde{B}_{124}$, and
$\widetilde{B}_{156}$.  (The other nonzero elements are obtained
by applying cyclic permutations $x\rightarrow y\rightarrow z$
or interchanges such as $x\leftrightarrow y$ on the Cartesian axes, e.g.,
$\widetilde{B}_{114}=\widetilde{B}_{225}$,
$\widetilde{B}_{124}=\widetilde{B}_{235}=\widetilde{B}_{215}$, and
$\widetilde{B}_{156}=\widetilde{B}_{225}$.)
We carried out {\it ab-initio} calculations of these tensor elements
using a plane-wave pseudopotential approach to DFT in the local-density
approximation (LDA) as implemented
in the ABINIT code package\cite{abinit}.
First, we relaxed the lattice
parameters for both GaAs and InAs.  
Next, linear-response calculations of the linear bulk piezoelectricconstant 
$\tilde{e}^0_{14}$ were carried out on these relaxed 
structures using the ANADDB module of the ABINIT package
\cite{wu05,abinit}, which implements a direct calculation of the
strain derivatives of the quantities of interest (Kohn-Sham
wavefunctions, polarizations, etc.) via the chain rule.
Then, a finite-difference
technique was used in order to obtain the non-linear bulk
piezoelectric tensors $\widetilde{B}_{\mu jk}$.  Specifically, 
we considered strain states of the
form $\eta_1=\eta_2=\eta_3=0$ and $\eta_4=\eta_5=\eta_6=\gamma$ for
several small values of $\gamma$.  With the strain frozen in for a
particular value of $\gamma$, the ions were allowed to relax, after which
the reduced piezoelectric tensor elements $\tilde{e}_{11}(\gamma)$,
$\tilde{e}_{12}(\gamma)$ and $\tilde{e}_{15}(\gamma)$
were computed using linear-response techniques as before.
The dependence of these elements on $\gamma$ was then fitted, and
the linear dependence extracted.  
From Eq.~(\ref{eq:estrain}),
this determines the three independent elements of the $\widetilde{B}$
tensor as
$\widetilde{B}_{114}=d\tilde{e}_{11}/d\gamma$,
$\widetilde{B}_{124}=d\tilde{e}_{12}/d\gamma$, and
$\widetilde{B}_{156}=d\tilde{e}_{15}/d\gamma$.
The results for GaAs and InAs are given in Table~\ref{tab:pcoef}.

\begin{table}
\caption{\label{tab:pcoef} Linear and quadratic piezoelectric
coefficients (C/m$^2$) as calculated from DFT.}
\begin{ruledtabular}
\begin{tabular}{lcccc}
     & $e_{14}$ & $B_{114}$ & $B_{124}$ & $B_{156}$ \cr \hline
InAs &  $-$0.115  & $-$0.531    & $-$4.076    & $-$0.120 \cr
GaAs &  $-$0.230  & $-$0.439    & $-$3.765    & $-$0.492 \cr
\end{tabular}
\end{ruledtabular}
\end{table}

{\it Calculation of the piezoelectric field for large structures 
(non self-consistent Poisson approach):} 
The DFT method cannot be applied to 10$^3$-10$^6$-atom structures
which are often of interest in nanoscience. 
Instead, such structures can only be calculated by 
non self-consistent methods (e.g., tight binding, $k.p$, empirical 
pseudopotentials), in which case piezoelectricity must be added 
as an external potential.
Thus, to model such structures, we first calculate
the strain tensor $\eta$ at each atom site using the valence force field (VFF)
method. In this method, the bond-stretching, bond-bending, and
mixed bending-stretching terms are derived from experiment.  The
method has been shown to give
accurate atomic positions for defect-free bulk and
alloys.
For example, Bernard and Zunger compared strain values obtained
by LDA and by VFF for the extreme case of a single monolayer InAs
superlattice in GaAs and obtaining agreement within 0.4\% \cite{bernard94}.
From a knowledge of the strain field $\eta_j(\vec{r})$, we can
use Eq.~(\ref{eq:red_pol}) to obtain $p_\mu(\vec{r})$, and
the piezoelectric charge density (per unit undeformed volume) is
then calculated from the divergence (in undeformed coordinates) of
$p_\mu$ via
\begin{equation}
\label{eq:rho}
\rho_{\rm piezo}(\vec{r}) = -\frac{e}{a_0^2}\nabla\cdot \vec{p}
\quad .
\end{equation}
The calculation of
the divergence [Eq. (\ref{eq:rho})] is performed using a piecewise
polynomial function to represent the polarization data
points. 
Finally, the piezoelectric potential
$V_{\rm piezo}$ is obtained from a finite-grid solver of the Poisson
equation
\begin{equation}
\label{eq:poisson}
\rho_{\rm piezo}(\vec{r}) = \epsilon_0\nabla\cdot
\left\{ 
\epsilon_s(\vec{r})\nabla V_{\rm piezo}(\vec{r}) \right\}
\quad ,
\end{equation} 
where we assume an isotropic local static dielectric constant $\epsilon_s(\vec{r})$.
However, note that the local polarization
cannot be defined on an arbitrarily small region of
space\cite{resta94}, but only on a scale that exceeds the
localization of the maximally localized Wannier functions
\cite{marzari97}.  For GaAs and InAs we
average the strain tensor over eight-atom clusters.
The piezoelectric tensor of Eq.~(\ref{eq:estrain})
is position dependent since it depends on the inhomogeneous strain.
Furthermore,  for alloys or heterostructures, $\tilde{e}_{\mu j} ({\bf r})$ 
also depends
on the material concentration at ${\bf r}$. In our case we
have regions in the cell with InAs, GaAs, or mixed (In,Ga)As, and we
use a linear interpolation
\begin{equation}
\label{eq:weighted_ave}
A({\bf r}) = x A_{\rm InAs}({\bf r}) + (1-x) A_{\rm GaAs}({\bf r})
\end{equation}
of the tensors ($A=e_{\mu j}$ or $B_{\mu jk}$) between the
constituent bulk materials for the given local concentration $x$ of the
eight-atom cell.
Finally, the solution of the Poisson
equation [Eq. (\ref{eq:poisson})] is obtained on the eight-atom-cluster
grid through a conjugate-gradient algorithm with a position-dependent
dielectric constant $\epsilon_s({\bf r})$ calculated
according to Eq. (\ref{eq:weighted_ave}) with $A=\epsilon_s$.
The approach developed for the calculation of $V_{\rm piezo}$ 
can be applied easily to very large 
(10$^6$-atom) nanostructures.

% Results

\begin{figure}
\includegraphics[width=.80\linewidth]{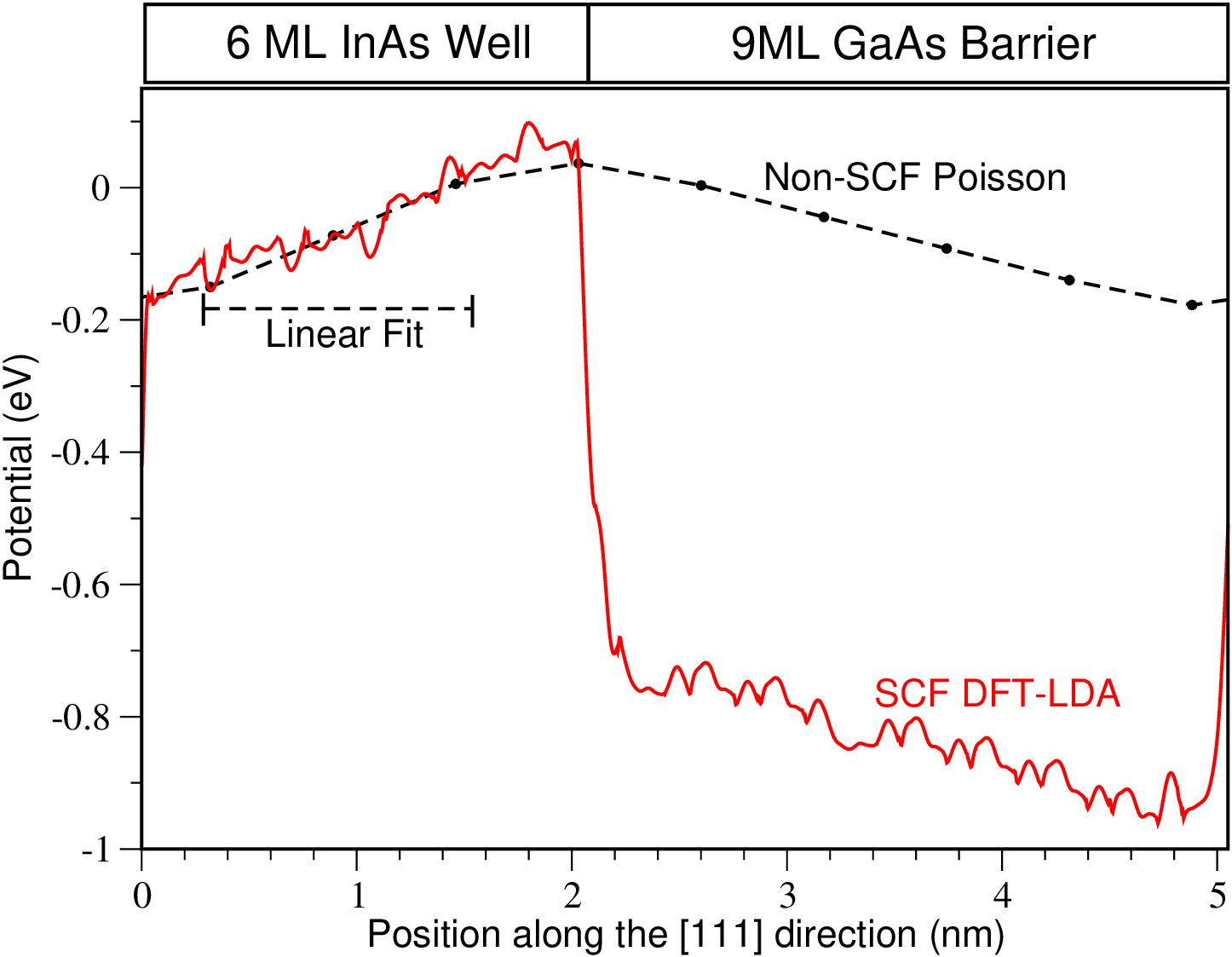}
\caption{(Color online) Piezoelectric potential calculated from
Eqs. (\ref{eq:red_pol}) to (\ref{eq:poisson}) (dashed line) and the
Kohn-Sham potential obtained from self-consistent DFT calculations
(solid line). Both calculations are for an (InAs)$_6$/(GaAs)$_9$ 
superlattice epitaxially strained on GaAs.\vspace{-0.1cm}}
\label{fig:DFTcomparison}
\end{figure}

% Fig. 1: Check with LDA
% Fig. 2: Potential vs [111] direction, different x. It is ~linear within the QW, we use 10,000 atoms.
% Fig. 3: Field vs. composition a) b), compare e=lda,b=lda with e=lda, b=0 (this is to show that
%just linear is far off), e=exp (this is a good
% fit to ~20 %) + comparison with exp.
{\it Testing the non-self-consistent Poisson approach:}
We tested the
non-self-consistent procedure described in Eqs.~(\ref{eq:red_pol})-(\ref{eq:poisson}) 
by comparing the results with direct
self-consistent DFT calculations for a model
quantum well of artificially small dimensions which can be handled by DFT. 
The system is a 30-atom [111] InAs quantum-well [(InAs)$_6$(GaAs)$_9$]
epitaxially strained to the GaAs in-plane lattice constant.
For the DFT calculations, we
used the same pseudopotentials and convergence parameters as in
the calculation of the piezoelectric tensors.  
To obtain the
piezoelectric field directly from the DFT-LDA quantum-well
calculations, we averaged out the atomic
oscillations from the self-consistent Kohn-Sham
potential (including ionic, Hartree and LDA
exchange-correlation contributions). The direct DFT results of this 
averaging
procedure\cite{baldereschi88} are denoted
in Fig.~\ref{fig:DFTcomparison} as the ``SCF DFT-LDA'' curve. (A linear
regression of the obtained curve in the region marked with ``Linear
Fit" gives a value for the piezoelectric field of 1255 kV/cm.) The
piezoelectric potential obtained with the procedure of 
Eqs.~(\ref{eq:red_pol})-(\ref{eq:poisson}) is given in
Fig.~\ref{fig:DFTcomparison} as the dashed curve denoted ``non-SCF 
Poisson".  The potential
jump at the interface is related to the band offset between
materials \cite{baldereschi88} and is not present in the bare
piezoelectric potential given by the dashed curve. 
The potentials are arbitrarily shifted to
coincide in the InAs region. We see that the field deduced from the 
non-SCF procedure is 1367 kV/cm, in very good agreement with the
self-consistent DFT-LDA quantum-well result of 1255 kV/cm.
  
\begin{figure}
\includegraphics[width=.80\linewidth]{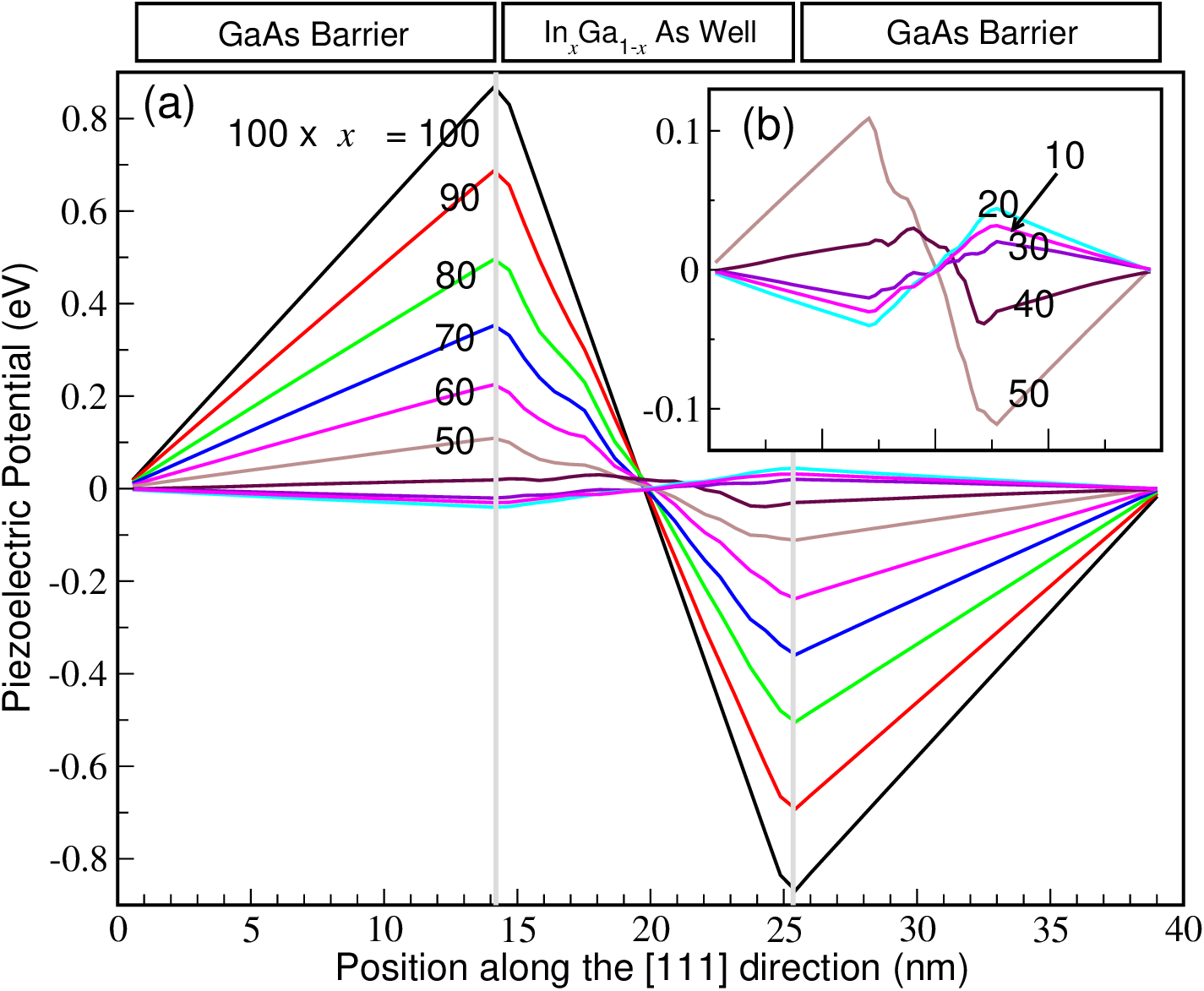}
\caption{(Color online) (a) Piezoelectric potential calculated from
Eqs. (\ref{eq:red_pol}-\ref{eq:poisson}) for an 11 nm
In$_x$Ga$_{1-x}$As well epitaxially strained to GaAs for
$x$ = 0.1 - 1.0. (b) for $x$ = 0.1 - 0.5.\vspace{-0.3cm}}
\label{fig:potential}
\end{figure}

{\it Application to large quantum-well structures.}
Having established the validity of the non-self-consistent method
for small systems, where comparison  with DFT-LDA is possible, we
now address nanostructures with sizes and composition typical of
experimental conditions. We chose here the quantum well system
In$_x$Ga$_{1-x}$As/GaAs where the  In$_x$Ga$_{1-x}$As alloy is
epitaxially grown on the GaAs substrate and the thickness of the
well is around 10 nm. 
We use 12,000 atoms in the simulation cell to accurately 
represent the random alloy.
We plot in Fig.~\ref{fig:potential} the piezoelectric
potential (in mV) along the [111] direction for In concentration
$x$=0.1 to 1.0. Figure \ref{fig:potential} shows a linear potential along the
growth axis, as expected from the fact that the piezoelectric
charges are well localized at the interfaces. The small
oscillations in the potential are due to random alloy fluctuations
which are most prominent at low In concentrations.  Interestingly,
Fig.~\ref{fig:potential} shows that the field reverses sign between
$x$ = 0.4 and 0.3, going from a very strong positive field
(i.e., negative slope along [111]) for In-rich wells to a weak negative
field in the In-poor regime.

\begin{figure}
\includegraphics[width=.85\linewidth]{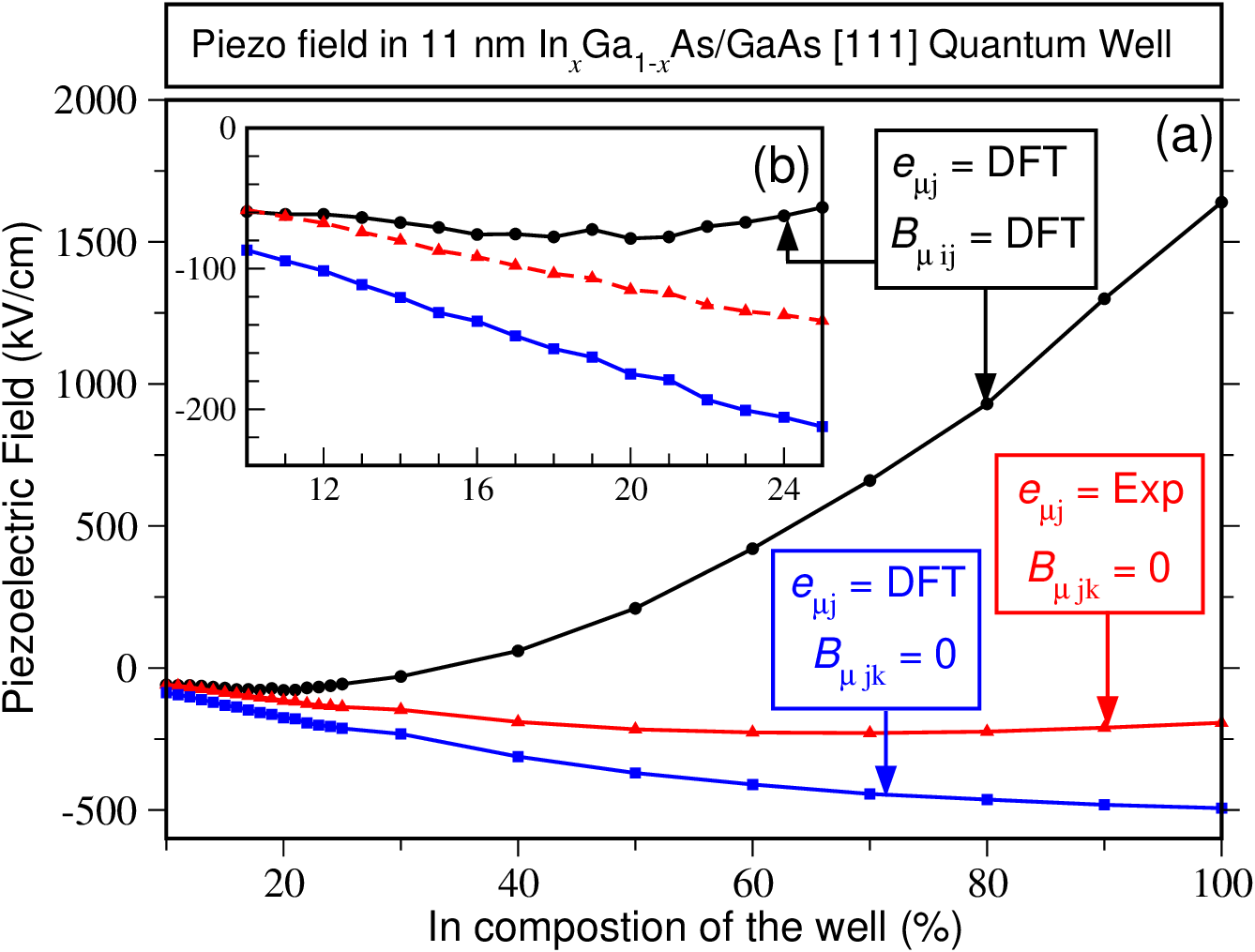}
\caption{(Color online) (a) Piezoelectric field as a function of $x$.
Circles: correct result with linear and
non-linear piezoelectric tensors from DFT. Triangles: neglecting
$B_{\mu jk}$ and using experimental $e_{\mu j}$. Squares:
neglecting $B_{\mu jk}$ and using LDA $e_{\mu j}$. 
(b) Magnification of the low-concentration region.\vspace{-0.3cm}}
\label{fig:field}
\end{figure}

The piezoelectric field extracted from
the potential of Fig.~\ref{fig:potential} is shown in 
Fig.~\ref{fig:field} as a
function of the In concentration (circles).  Figure~\ref{fig:field}(b) 
shows the electric field with an emphasis on
the experimentally relevant concentration range from $x$= 0.10 to
0.25. We see that the amplitude of the field obtained with both linear and nonlinear 
terms is much smaller than the field obtained with $e_{\mu j}$ only. Considering 
the full concentration range, the field obtained using both $e_{\mu j}$ and $B_{\mu jk}$ is
shown to be negative at low In concentrations and almost constant until it
reaches 30\%, where it reverses sign and becomes very strong. The field we
obtain for a concentration range of 16-20\% In is nearly constant around 80 kV/cm.

A direct comparison of calculated and measured electric fields is difficult because
only few experiments report the value of the measured field and the measurements
and calculations are performed on different concentrations and thicknesses. 
Cho {\it et al.}\cite{cho01} obtained a field of 129 $\pm$ 12 kV/cm for a  17 \% In well with an
estimated 8.7 nm thickness, and Sanchez {\it et al.} obtained a field
of 137 $\pm$ 6 kV/cm for a 17\% In well with 10 nm thickness and  121 $\pm$ 5
kV/cm for a 10 nm thick well with 21\% In concentration. Furthermore, the results are 
clouded by strong temperature effects \cite{bahder94,cho01,sanchez02} (pyroelectricity)
and possible effects of In segregation \cite{ballet99}.
However, one experimental observation that does not seem to depend on well
concentration and thickness is the fact that the measured piezoelectric 
field leads to an $e_{14}$ value
that is about 35\% smaller than what is expected by using the linear 
coefficients alone. This has been reported on many occasions 
\cite{bahder94,sanchez-rojas94,chan98,ballet98,cho01} and constitutes an
unsolved puzzle.
This result can be accurately compared with our calculations. If we calculate the 
piezoelectric field using the experimental value of $e_{14}$ for bulk InAs ($-$0.045
C/m$^2$) and bulk GaAs ($-$0.160 C/m$^2$) and neglecting the 
second-order tensors $B_{\mu jk}$ in Eq. (\ref{eq:red_pol}) 
(triangular symbols in Fig.~\ref{fig:field}), which is equivalent to the experimental 
procedure that leads to an overestimation of $e_{14}$ by 35\%, we find 
for the concentration region of $\simeq$18-21\%, an overestimation of the magnitude of the
field by 34-52\%.
Our results therefore explain the
origin of the experimentally observed deviation
\cite{bahder94,sanchez-rojas94,chan98,ballet98,cho01}
: a linear interpolation between the InAs and GaAs values of
$e_{14}$ cannot reproduce the piezoelectric field of alloyed
quantum wells since the field does not originate from the linear
coefficient alone but has significant contributions from the 
second-order piezoelectric tensors
$B_{\mu jk}$ (neglected in the analysis of the experimental results).

To emphasize the effect of the non-linear tensors $B_{\mu jk}$
further, we plot using square symbols in Fig.~\ref{fig:field} the
piezoelectric field with $e_{\mu j}$ set to the DFT values and $B_{\mu
ij}$ set to zero.  The results show that, when taking only the
linear tensor into account, the field is overestimated by about
200\% in the region of low concentration [Fig. 3(b)], and even has
the wrong sign at higher concentrations.

% CONCLUSION
In summary, we have shown that the second-order piezoelectric
tensor, generally neglected so far in theoretical and experimental work,
contributes significantly to the piezoelectric effect in
zincblende semiconductors. We showed
that the piezoelectric field calculated by including first- and
second-order piezoelectric tensors obtained from DFT agree
well with experiments, whereas neglect of non-linearities 
leads to qualitative disagreements. We argue that the
``piezoelectric coefficients'' that have been extracted from
experimental work so far are actually effective ones reflecting
equally strong first- and second-order contributions.

This work has been supported by U.S. DOE-SC-BES-DMS under LAB 03-17
initiative and by NSF Grant DMR-0233925.

%\bibliography{DotBiblio,books}% Produces the bibliography via BibTeX.

%Figure 1

%Figure 2

%Figure 3

\end{document}